# Assign Hysteresis Parameter For Ericsson BTS Power Saving Algorithm Using Unsupervised Learning

Thaer Sahmoud[1], Wesam Ashor[2]
[1]Thaer.sahmoud@students.iugaza.edu.ps, [2]Washour@iugaza.edu.ps
[1,2]Computer Engineering Department, Islamic University of Gaza, Palestine

*Abstract- Gaza Strip suffers from a chronic electricity deficit that affects all industries including the telecommunication field, so there is a need to optimize and reduce power consumption of the telecommunication equipment. In this paper we propose a new model that helps GSM radio frequency engineers to choose the optimal value of hysteresis parameter for Ericsson BTS power saving algorithm which aims to switch OFF unused frequency channels, our model is based on unsupervised machine learning clustering K-means algorithm. By using our model with BTS power saving algorithm we reduce number of active TRX by 20.9%.*

## 1. INTRODUCTION

Mobile communications technology is often classified to several generations, the First Generation "1G" was an analog system designed to carry voice only, after that the "1G" was replaced by a digital communication system called the Second Generation "2G" which is capable of carrying both voice and data traffic. Later, the Third Generation "3G" **[1]** Fourth Generation "4G" **[2]** and Fifth Generation "5G" **[3]** networks can supply higher data-rate radio access networks. In Gaza Strip, telecommunications operators are allowed to use only 2G networks.

The Global system for mobile communication "GSM" **[4]** is a cellular-wireless telecommunications standard developed by the European Telecommunications Standards Institute, it is a "2G" network technology and originally was developed to transmit voice, but it is also able to transmit data via circuit-switched network. Packet data capabilities were added to the GSM slandered by means of General Packet Radio Services "GPRS" **[5]**, the next advance in GSM radio access technology was Enhanced Data rates for Global Evolution "EDGE" **[6]**- also known as Enhanced GRPS, EDGE is actually the mobile wireless network that is deployed in Gaza Strip.

GSM Radio Access Network "RAN" consists of a number of Base Transceiver Station "BTS", each BTS divided into a number of cells and to each cell we assign a number of frequencies called Transceivers "TRX" for data communication between user mobile station "MS" and the GSM network, cells TRXs number depends on traffic load (voice and data) such that high loaded cells assigned a greater number of TRXs than low traffic loaded cells, and GSM use multiple access technique called Time Division Multiple Access "TDMA" **[7]**, so that each TRX is divided into 8 time slots "TS" called physical channels. Each channel can carry one full rate voice call or two half rated voice calls.

Gaza Strip has a critical electricity crisis **[8]**, since in the best case, residents usually receive power in eight-hour rotations: eight hours ON and eight hours OFF, and in summer, the power can go OFF for up to 12 hours or even 16 hours.

While electricity OFF, BTS will run on battery bank, and when battery voltage crosses a threshold, then a power generator - which exists in BTS site- will run to charge the batteries and supply the BTS with the needed power. So, reducing BTSs power consumption leads to increase Battery life, reduce power generator fuel consumption, reduce heat produced by equipment and reduce the overall network operation cost.

In this paper we propose a new model that enable us to set an optimal hysteresis parameter value for Ericsson BTS power saving using K-means clustering algorithm **[9]**. We first use K-means to group GSM network cells into three clusters according to cells traffic readings, after that we utilize the clustering result to assign the BTS power saving parameter for each cluster so that we can reduce BTS power consumption by turning OFF unused TRXs, while maintaining almost the same network performance.

## 2. RELATED WORK

Many researches discuss telecommunication power saving techniques, so that network operators can reduce network operation cost by optimizing power consumption. **[10]**, **[11]** use Monte Carlo simulation algorithm to select the best values for BTS power saving parameters, **[12]**implement unsupervised algorithm to analyze 4G cells behavior from uplink performance perspective and group network cells into clusters that have commonalities in uplink behavior and the implementation of that algorithm results an improvement in uplink speed for live 4G network by 7%. **[13]** addressed the key challenges of envisioning the hybrid solar powered BTSs in Bangladesh considering the dynamic profile of the renewable energy sources and traffic intensity and proposed system to downsize the electricity generation cost and waste outflows while ensuring the desired quality of experience "QoE" over 20 years duration. **[14]** proposed an approach for optimizing GSM power control algorithm to enhance GSM voice performance, while **[15]** presented a new model to study energy saving strategies in the telecommunication apparatuses of Base Transceiver Stations "BTSs" using Monte Carlo simulation of power saving function, **[16]** proposed a modification to the existing rooftop BTS located in urban area by adding a small solar photo voltaic power plant to existing rooftop BTS, **[17]** presented trial results and network energy saving estimations when several energy saving software based features are activated for different network site configurations, **[18]** analyze the impact of turning off LTE radio without moving low cost machine type communication into Radio Resource Control "RRC" Idle and reducing the RRC Connected tail time, **[19]** proposed a new channel, forward wake-up channel "F-WUCH" for CDMA2000 1xEV-DV (evolution data and voice) systems , to maximize the power saving of the mobile station, **[20]** discussed Discontinuous Reception mechanism "DRX" parameters and features in TD-LTE

In section 3 we explain in brief details Ericsson BTS power saving algorithm and how it works, and in section 4 we describe our proposed model that we use to assign the suitable hysteresis parameter for BTS power saving algorithm, and then the experimental results are in section 5.

### 3. BTS POWER SAVING ALGORITHM

There are many types of information that have to be transmitted between the user MS and BTS, this information is transmitted on logical channels. Logical channels can be divided into two types, control channels and traffic channels "TCH", control channels "CCHs" are dedicated to the sending and/or receiving of command messages between BTS and the user MS, it contains main information about the cell like cell ID, frequencies that assigned to that cell so the user can use for communication and other information required for signaling, while traffic channels are channels that carry traffic voice and data, for voice TCH there are full rate traffic channel "TCH/F", that TDMA time slot is allocated for one voice call, and there is also half rated traffic channel "TCH/H" that two "TCH/H" can allocated in one channel, while TCH for packet data is Packet Data Traffic Channel "PDCH".

The method of placing logical channels on physical channels is called mapping, while most logical channels take one physical channel to transmit, some take more. Control channels are usually assigned to the first two or three physical channels in the first TRX for each cell, and the rest of physical channels are assigned as traffic channels, voice traffic is measured in Erlang, in practice, it is used to describe the total traffic volume of one hour (e.g., one Erlang equals one call with call duration equals one hour, or equals also two calls with sum of call durations one hour for the two call durations, and also equals six calls with call duration of ten minutes each, and so on).

Traffic load in GSM network varies depending on the geographical location of the BTS, so that BTS that serve -for example- universities will have high traffic on student attendance time interval and will have a low traffic after student attendance interval or on vacations, also, traffic load varies depending on time of day, and the day of the week or year, as illustrated in figure 1 which shows traffic load for a cell for 6-days interval, and from that figure it is clear that the day time has more traffic than night time where traffic may almost down to zero.

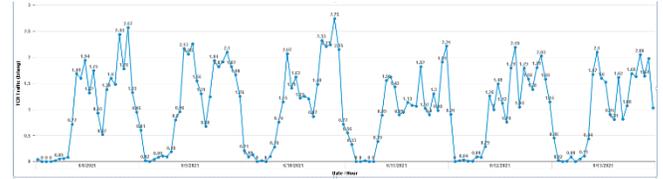

Figure 1: Traffic profile for a sample cell for 6 days

To reduce power consumption on BTS, we use BTS power saving, which is an optional feature in Ericsson BTS that save power by turning off unused TRXs, noting that since the first TRX contains "CCHs" which as mentioned earlier are responsible for signaling between the cell and user "MS" so we cannot turn OFF that TRX because turning it OFF means the whole cell will be down and unseen from user "MSs". The decision to turn off a TRX is done by measurement of the following parameters:

- TRXOFFTARGET: this parameter specifies the number of intervals with traffic under the low-traffic threshold level before disabling a TRX with BTS power saving.
- TRXONTARGET: this parameter specifies the number of intervals with traffic over the high-traffic threshold level before enabling a TRX with BTS power saving.
- TRXOFFDELAY: this parameter specifies the interval between disabling a TRX within a cell due to BTS Power Savings and the resumption of checking for a further TRX to disable for BTS Power Savings in the same cell. The value is given as a number of scans, where the time between each scan is approximately 10 seconds.
- BTSPSHYST: This parameter changes the hysteresis for BTS Power Savings. The value is expressed as the number of idle TCH.

Table 1 shows range value for these four parameters and the Ericsson recommended value for each.

TABLE 1: BTS POWER SAVING PARAMETERS VALUES

| Parameter | Range Value | Default value |
|---|---|---|
| TRXOFFTARGET | 20 to 100 | 50 |
| TRXONTARGET | 20 to 100 | 49 |
| TRXOFFDELAY | 6 to 90 | 30 |
| BTSPSHYST | 1 to 1014 | 5 |

A TRX is turned OFF if it remains unused for a certain time, BTS use cyclic power saving algorithm developed by Ericsson **[21]** as shown in figure 2 that runs every 10 seconds, the algorithm has an initial counter "N =0", and fixed offset equals 9, the algorithm checks every 10 seconds if the number of idle time slots is greater than the sum of BTSPSHYST and the fixed offset, if that condition is true then the counter "N" is increased by one, else the counter "N" is the max of (N-3, 0), at the time when "N" equals

TRXOFFTARGET then a TRX is turned OFF, and after turning off a TRX, the algorithm wait for time equals to TRXOFFDELAY and after that reset the counter "N" and do the same loop.

On the other hand, a TRX is switched ON when the counter reaches the value TRXONTARGET. This counter is increased only when the number of idle time slots is less than the summation of BTSPSHYST and the fixed offset.

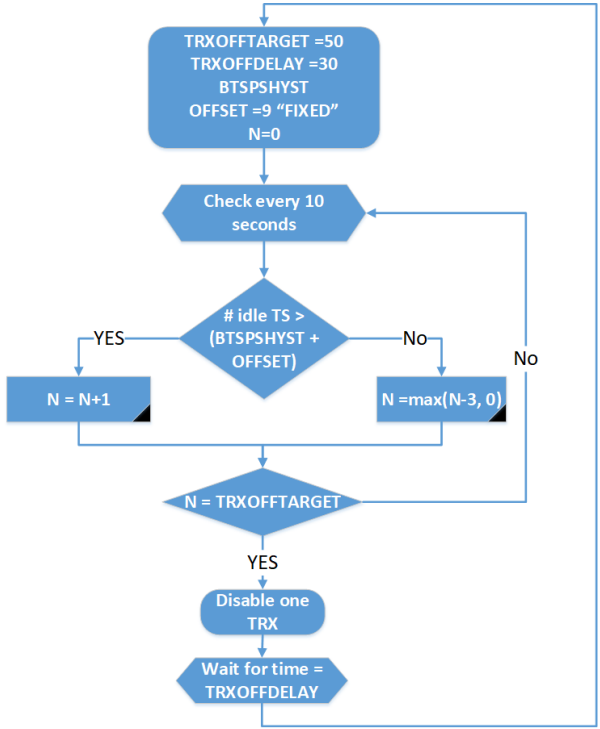

Figure 2: Ericsson BTS Power Saving Algorithm

A high hysteresis "BTSPSHYST" ensures a high number of Idle BPC when a TRX is disabled. As a result, decreases the risk of congestion because of fast changes of traffic load in the cell and also increase power consumption. While, a low hysteresis gives more energy savings but increase the chance of congestion occur at the cell due to fast increase in traffic. So that, the hysteresis "BTSPSHYST" parameter must be balanced between optimizing energy savings and preserving sufficient signaling and packet switched traffic capacity during low traffic periods. In our paper, we implement an unsupervised clustering algorithm that cluster the cells in our network into three clusters based on traffic parameters, and after that we assign a BTSPSHYST value for each cluster that balanced energy saving and network performance.

when we set BTS power saving OFF, TCH mapping onto physical channel does not happen in order mode, that means a new call does not take the first free physical channel, it can take any free physical channel based on time slot quality, figure 3 shows an example for mapping logical channels onto physical channels for a cell with three TRXs, and from that figure we can see that we have only nine active physical channels while the three TRXs are active and consume power.

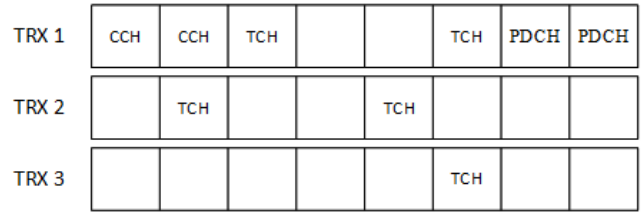

Figure 3: Mapping logical channels onto physical channels without BTS power saving.

on the other hand, when we enable BTS power saving and using a suitable BTSPSHYST value, for the same above example cell, TCHs will occupy the first and second TRXs as shown in figure 4, and let the third TRX unused and thus can be switched OFF.

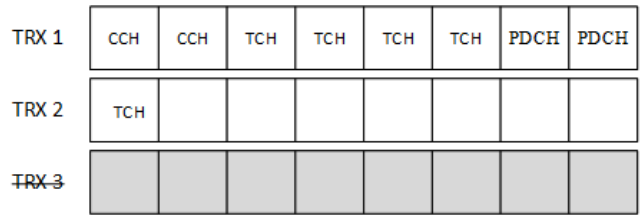

Figure 4: Mapping logical channels onto physical channels with BTS power saving.

## 4. THE PROPOSED MODEL

As we see in the BTS power saving algorithm, hysteresis parameter "BTSPSHYST" is the most critical parameter that we need to care of, and since low "BTSPSHYST" value results to decrease number of idle TRXs an so low power consumption but on the other hand low "BTSPSHYST" leads to traffic congestion due to any surge in traffic. So, we need to set "BTSPSHYST" based on each BTS traffic load.

One simple way to assign a suitable value for "BTSPSHYST" is to check BTS traffic (voice and data) and all related readings and after study the traffic behavior for that BTS we can assign "BTSPSHYST", but it will take a lot of time to do such a process specially when we have a large amount of cells in the network and when traffic change from time to time, so we propose our model that aim to group BTS cells into number of clusters based on traffic load – and all traffic related readings- for each cell, so that similar cells in traffic grouped in the same cluster and then we assign one "BTSPSHYST" value for all cells belongs to the cluster. By applying our model, we can assign "BTSPSHYST" value for all cells in the network in a few minutes, so we can follow up and update these values from day to day or when required.

Our model in based on K-means clustering algorithm which aim to partitioning cells into "K" clusters based on cells traffic readings. Before applying the K-means clustering algorithm, we need to know the optimal number of clusters that we can divide our data set into. To do so, we use the elbow method which is one of the most popular methods to determine this optimal value of the number of clusters "K". Elbow method runs K-means clustering algorithm to the data set for a range of clusters (e.g., k =1 to 10) and for each cluster it computes the Sum of Square Error "SSE" and plot a graph between "SSE" and "K", the location of the bend is the considered as the suitable number of classes. Figure 5 shows the elbow method for our dataset, we can see that the bending occurred at K=3 and also at K=6, so we use the Silhouette Coefficient [23] to make sure which value of K is the optimal one.

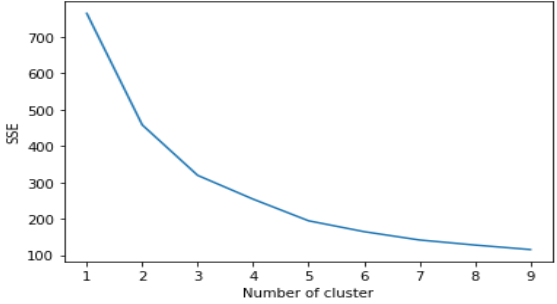

Figure 5: Elbow graph for our dataset

Silhouette Coefficient is a metric used to calculate the goodness of a clustering technique. Its value ranges from -1 to 1 and can be calculated using the below formula (1)

$$Silhouette\ Score = \frac{(A-B)}{max\ (A,B)} \quad (1)$$

Where "A" is the average intra-cluster distance and "B" is the average inter-cluster distance, when silhouette score is "1" then we can say that clusters are well separated, while if silhouette score equals "0" that means the clusters are indifferent, and when the score is "-1" then we say that clusters are assigned in incorrect way. To use silhouette score method, implement K-means for range of clusters (K=2 to 9) and for each value of "K" we calculate silhouette score and plot silhouette score vs number of clusters "K" Figure 6 shows silhouette for our dataset, and since the highest score occurred when number of clusters equals "3" then "K=3" is the optimal number of clusters for our data set.

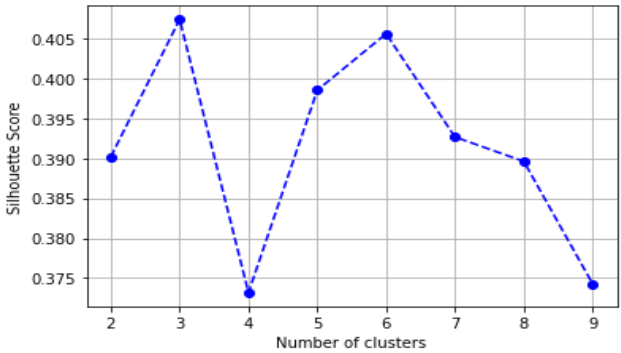

Figure 6: Silhouette score for our dataset

K-means [9] is a partitional clustering algorithm, the main idea is to select "K" centroids -also called seeds- randomly as initial centroids, then assigning all data points to the closest centroid so that each centroid forms a cluster, after that we recompute the centroid for each cluster and repeat until the centroids for each cluster do not change. One of the disadvantages of K-means is that it is sensitive to the initial centroids location which means that different initial centroids locations may leads to different clustering result, to overcome that issue we use K-means++ algorithm [24] which choose the initial centroids by firstly select randomly one point ($x$) from data points ($X$) as the first centroid ($c_1$), then next centroid ($c_2$) is chosen as ($x' \in X$) such that the probability function (2) is maximize, where $D^*(x)$ is the shortest distance between a datapoint $x$ and its closest centroid already chosen.

$$\frac{D^*(x')^2}{\sum_{x \in X} D^*(x)^2} \quad (2)$$

After choosing the second centroid we repeat choosing the next centroid using the same method that the second one has chosen. We repeat that method until all "K" centroids are chosen, once all centroids are chosen K-means++ works in the same manner just like the original K-means. we use python "Scikit-learn" library to implement K-means++ with number of clusters K=3. figure 7 shows K-means result, and we can see that the three clusters have some separation between each other.

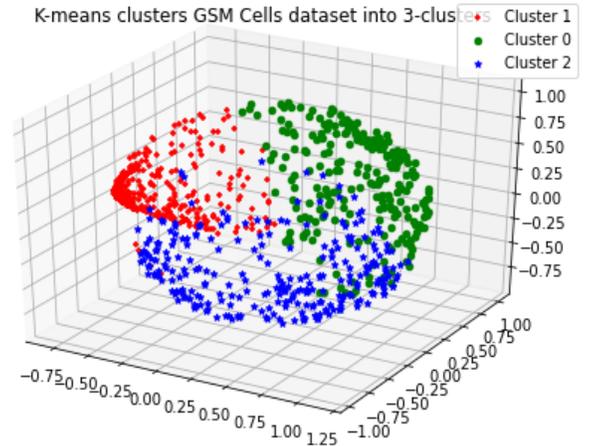

Figure 7: K-means Clusters

## 5. EXPERIMENTAL RESULTS

### 5.1. DATASET AND PARAMETER SELECTION

Since BTS power saving affects BTS traffic both voice and data, so we collect traffic readings for cells from 1-January-2021 to 30-March-2021, we take daily busy hour readings and average these ready for each cell, so that if there is outlier data point due to impulse traffic will be normalized. The following Key Performance Indicators "KPIs" are taken in account while clustering cells as data point features:

- TCH Traffic (Erlang): Total traffic for cells in Erlang.

- PREEMPTPDCH: Total number of used PDCHs that has been pre-empted by voice traffic, because of voice channel congestion.
- PDCH Congestion: the percentage of data packet congestion.
- DL EDGE Throughput: throughput for downlink from BTS to user MS data packet in Kbps.
- Number of TCHs: number of time slots to carry traffic voice and data.

Table 2 shows data points sample, and since data features have different scales and that results the feature with higher value will dominant other small value features, so we do a preprocessing step by normalizing and standardizing the dataset so that all features will have a similar weight after that we reduce features number to three features using Principal Component Analysis "PCA" **[22]**.

TABLE 2: DATASET SAMPLE

| Cell ID | TCH Traffic Erlang | DL EDGE Throughput (Kb/s) | PDCH Congestion (%) | PREEMPT PDCH | TS |
|---|---|---|---|---|---|
| Cell_1 | 2.69845 | 130.523 | 0.00579 | 5.08791 | 24 |
| Cell_2 | 1.62493 | 136.034 | 0.00596 | 3.12088 | 24 |
| Cell_3 | 7.31606 | 124.882 | 0.11292 | 41.95604 | 32 |
| Cell_4 | 5.25773 | 123.006 | 0.01373 | 16 | 32 |
| Cell_5 | 4.42022 | 132.727 | 0.00066 | 2.04396 | 24 |

### 5.2. RESULTS

After clustering network cells into three clusters, we assign a suitable hysteresis "BTSPSHYST" value for each cluster by taking samples cells from each cluster and review readings for these sample cells after that we can give each cluster the suitable BTSPSHYST parameter value, below table 3 shows a sample for K-means clustering result.

TABLE 3: SAMPLE OF K-MEANS CLUSTERING RESULT

| Cell ID | TCH Traffic Erlang | DL EDGE Throughput (Kb/s) | PDCH Congestion (%) | PREEMPT PDCH | TS | Cluster |
|---|---|---|---|---|---|---|
| Cell_1 | 2.69845 | 130.523 | 0.00579 | 5.08791 | 24 | 1 |
| Cell_2 | 1.62493 | 136.034 | 0.00596 | 3.12088 | 24 | 1 |
| Cell_3 | 7.31606 | 124.882 | 0.11292 | 41.95604 | 32 | 2 |
| Cell_4 | 5.25773 | 123.006 | 0.01373 | 16 | 32 | 2 |
| Cell_5 | 4.42022 | 132.727 | 0.00066 | 2.04396 | 24 | 0 |
| Cell_6 | 4.86402 | 139.305 | 0.00065 | 3.91209 | 24 | 0 |

Based on K-means result as in table 3, cluster "2" cells have high traffic in Erlang, and a below accepted range of packet throughput with high value of packet data preemption and packet data congestions, so in that case we assign a BTSPSHYST value equals 12 for all cells in that cluster, on the other hand, we can we that cluster "1" has low Erlang traffic and a higher packet data rates that in cluster "2" with data packet preemption and congestion, so BTSPSHYST value of 4 is suitable for that cluster, and finally for cluster "0" it has a medium Erlang traffic with high packet data rate and low packet preemption and congestion, so for cells in cluster "0" we assign a value 6 for BTSPSHYST parameter.

After assigning BTSPSHYST for each cluster, we collect traffic and number of average TCHs -which indicate number of active TRXs since every eight TCHs equals one TRX except for the first TRX where we have three CCHs so there is 5 TCHs for that TRX- figure 8 shows average traffic for a sample cell and number of TCHs before activating power saving, as we see in that figure number of active TCHs is 21 that means we have 24 time slot "21 TCHs and 3 CCHs" so that cell has three active TRXs despite the status of traffic. On the other hand, figure 9 shows traffic vs active TCHs for the same cell after applying BTW power saving, and we can see that number of active TCHs depends on traffic load on the cell, and the figure shows that max number of TCHs for that cell after activating BTS power saving is 13, so we have only two TRXs active and 1 TRX switched off and that number and number of active TRXs on night reaches only one active TRX (for active TCHs equals 5).

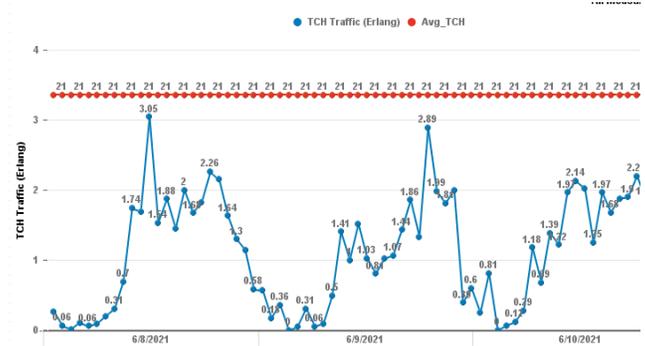

Figure 8: Traffic vs average TCHs for a cell without power saving

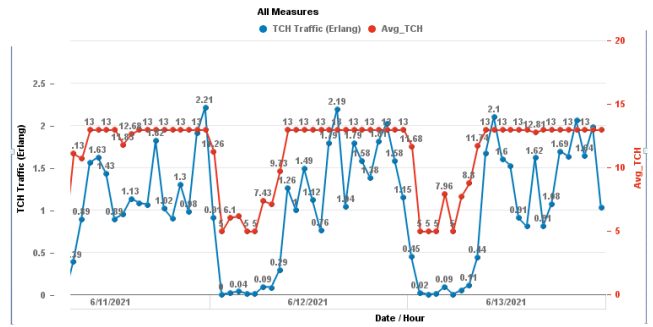

Figure 9: Traffic vs average TCHs for the same cell with power saving

Below table 4 shows number of active time slots per a sample group of cells before and after BTS power saving, and as illustrated on that table using BTW power saving reduces the number of active time slots for most cells in the network except for heavy loaded cells such as cell_11 and cell_12.

TABLE 4: NUMBER OF ACTIVE TIME SLOTS BEFORE AND AFTER BTS POWER SAVING

| Cell ID | TS before BTS power saving | Max num. of TS after BTS power saving |
|---|---|---|
| Cell_1 | 24 | 14 |
| Cell_2 | 24 | 13 |
| Cell_3 | 24 | 14 |
| Cell_4 | 24 | 13 |
| Cell_5 | 24 | 14 |
| Cell_6 | 24 | 14 |
| Cell_7 | 32 | 13 |
| Cell_8 | 24 | 16 |
| Cell_9 | 24 | 14 |
| Cell_10 | 24 | 18 |
| Cell_11 | 32 | 32 |
| Cell_12 | 32 | 32 |

So, as illustrated in figures 8,9 and table 4, clustering network cells into groups based on traffic load per cell using unsupervised machine learning enables us to assign suitable BTS power saving parameter "" for Ericsson power saving algorithm that results to reduce number of active TRXs which leads to reduce power consumption, heat dissipation, fuel consumption and the network operational cost and can reduce the impact of Gaza Strip electricity issue on our network.

## 6. CONCLUSION

Power consumption optimization is an important task for telecommunication operators so that operators can reduce the network running cost, and since Gaza Strip suffers from a critical electricity issue, then we in Gaza give the reduction of power consumption more priority. That is what motivate us to proposed our model which has been developed and tested on a GSM network that operates in Gaza, and results to reduce number of active TRX from 5754 (without activating BTS power saving) to 4550 (After activating BTS power saving using our proposed model) which means that we reduce TRX power consumption by 20.9%.

For future work, we are planning to include more related cells readings and implement our model for other BTS parameter assigning.